\documentclass{article}
\usepackage{authblk}
\usepackage{graphicx}
\usepackage{subfigure}
\usepackage{amsmath}
\begin{document}
\title{Ostrogradsky instability and Born-Infeld modified cosmology in Palatini formalism}
\author{Srivatsan Rajagopal\thanks{chichieinstein@sify.com}\\Ajit Kumar\thanks{ajitk@physics.iitd.ac.in}}
\affil{Department of Physics, Indian Institute of Technology\\
       Hauz Khas, New Delhi-110016}
\maketitle
\begin{abstract}
The Ostrogradsky instability of higher derivative Lagrangians is derived from first principles
using Control theory and Lyapunov Stability Analysis. This result is then used to argue that Born-Infeld
Lagrangians are viable modifications of the Einstein-Hilbert action provided the action is varied in
accordance with the Palatini formalism, in contrast to the metric formalism. Finally, the Born-Infeld version of
the FRW equations are derived and the cosmological dynamics is studied for matter dominated closed and
open universes, and the results are compared with the usual cosmology.
\end{abstract}
\section*{Introduction}
The light curves plotted from several hundred Type Ia supernovae [1] indicate that
the universe is expanding at an accelerated rate. On the other hand, the Friedmann
equations derived from Einstein's General Relativity show that for a fluid that sat-
isfies the Weak Energy Condition, the universe should decelerate as it expands [2].
This astonishing discrepancy between observation and the prediction of the general
theory of relativity led to a spurt of research activities intended to modify the existing
cosmological models with an aim to make them consistent with observation.
\\
\indent The endeavor started with models that include either the Cosmological Constant,
Quintessence, Dark Energy, or Phantom Energy [3] that presumably drives the accelerated expansion. However, all such models faced one or the other problems. For
instance, in the case of the models with cosmological constant, the smallness of its
value inferred from astronomical
data compared to its value predicted by particle physics and the remarkable but implausible coincidence of the vacuum energy density
and the mass density in the present epoch, require two-fold fine tuning which does not
appear appealing at all. On the other hand, models advocating for the existence of
dark energy suffer from their own problems related to the very nature of dark energy
and the fact that different models predict different rates of expansion that require a
rather tedious task of searching for the fingerprints of the dark energy in the entire
history of cosmic expansion. Added to it is the unpleasant fact that distinguishing
among different histories of expansion would require measurements an order of magnitude more accurate than the existing ones [12]. 
\\ 
\indent Apart from that, in order to make
Einstein's field equation with dark energy agree with the observed anisotropies in the
cosmic microwave background, the amount of dark matter that should be posited
is eighteen times the observed ordinary matter. The unresolved, so far, problems
related to these models inspired a third route, namely, the modification of the Einstein's field equations themselves by taking into account the higher order derivative
terms in the Einstein-Hilbert (E-H) action. 
\\
\indent However, as it turned out, this cannot
be accomplished in an offhand manner. There is a no-go theorem that constrains the
form of the Lagrangian that can be used for this purpose. In its essence, it is related to
Newton's observation that the equations of physics, when written in terms of the fundamental quantities, are always second-order in time [3]. This no-go theorem is called
Ostrogradsky's theorem: ``A system whose Lagrangian depends non-degenerately on
the second- and higher-order derivatives of the dynamical quantities is necessarily un-
stable''. The terminology \textit{non-degenerately} means that the higher derivatives can
be solved in terms of the lower derivatives.
This theorem has been invoked in the literature to support the claim that the
modifications of the E-H Lagrangian that depend on the traces of the Riemann and
Ricci tensors must be excluded from consideration [3]. But the justification of this
important result has been done by bringing ideas of second quantization into the
classical setting. Moreover, the sign indefiniteness of the energy only shows that the
energy function is not a Lyapunov function. It does not indicate that no other Lyapunov function can exist.

In the given paper, we first show that the proof of Ostrogradsky's theorem can be made more precise within classical mechanics, without discussing the particle picture. Then we proceed on to show that, if we consider a $f(R)$ theory of gravity by modifying the E-H Lagrangian with the Born-Infeld (B-I) type of nonlinearity, Ostrogradsky's theorem can be circumvented under the condition that the modified action is varied only in Palatini's formalism. In view of this, our work is devoted to the study of the B-I modified gravity model in Palatini's formalism. The cosmological implications of such a model are studied for the case when the spatial hyper-surface can be curved. Note that most of the works on B-I cosmology \cite{Ban2} have been restricted to the case of flat spatial hypersurfaces.\\ 

The organization of the paper is as follows. We start with the proof of Ostrogradsky's theorem in the classical mechanical framework. We then take up the modified E-H Lagrangian with B-I nonlinearity and justify why the corresponding action should be varied only in the Palatini's formalism. Finally, we derive Friedmann-Robertson-Walker (FRW) equations for the B-I gravity, study their cosmological implications and present the results of our numerical calculations

\section*{Chaetev's Theorem}
\indent We first review a theorem from control theory. This result is very important for the subsequent discussion with regard to the proof of the Ostrogradsky instability.\\
\indent Consider the following system of differential equations
\begin{align}
   \textbf{\.{x}} = f(\textbf{x})
\end{align}
Here, \textbf{x} represents an n-dimensional vector, while f(\textbf{x}) is a scalar point function. Without losing generality, we can assume that \textbf{x} = 0 is the equilibrium point of the system. We now state the result. The proof can be looked up in \cite{Kh}.

Chaetev's Theorem \cite{Kh} : Let $\mathbf{x} = 0$ be an equilibrium point of the system for (1). Let $V : D\longrightarrow R$ be a continously differentiable function such that $V(0) = 0$ and $V(\mathbf{x_0}) > 0$ for some $\mathbf{x_0}$ with $||\mathbf{x_0}||$ arbitrarily small. For $r>0$, let $B_r$ denote the set of all $\mathbf{x}$ with $||\mathbf{x}||$ less than $r$ contained in $D$ and let  
\begin{align}
 U = \left\{ \textbf{x} \in B_r | V(\textbf{x}) > 0 \right\}
\end{align}
 Additionally let $\dot{V} > 0$ throughout U. Then the point \textbf{x} = 0 is unstable. \\
\indent Here, the instability referred to, is instability in the Lyapunov sense. When translated in
terms of the particle picture, this amounts to exactly the same sense as conveyed in [3]. However, this result is more general and can be applied even to a classical field theory, like General Relativity, which cannot be quantized unambiguously at present. Now, we apply this result to the case of the system with a Lagrangian that depends non-degenerately on the second and higher order derivatives of the dynamical quantities. For simplicity and without loss of generality, we treat the case of a system with a finite number of degrees of freedom.

\section*{Ostrogradsky Instability and its consequences} 
\indent In \cite{Wood}, the Ostrogradsky Hamiltonian for one degree of freedom has been constructed explicitly for a higher derivative Lagrangian $L(q, \dot{q}, \ddot{q})$ to be of the following form
\begin{align}
H(q_1, q_2, p_1, p_2) = p_1 q_2 + \mu(q_1, q_2, p_2) - L (q_1, q_2, \mu)
\end{align}

Here, the choice of canonical coordinates is as follows \cite{Wood}
\begin{align}
q_1 &= q; \hspace{0.1 in} p_1 = \frac{\partial}{\partial \dot{q}} L - \frac{d}{dt}\frac{\partial L}{\partial \ddot{q}} \\
q_2 &= \ddot{q}; \hspace{0.1 in} p_2 = \frac{\partial}{\partial \ddot{q}} L
\end{align}
An approach to proving the instability of a system was already carried out in \cite{Akk}, in the
context of charged solitons. The ideas presented below are a combination of the steps
taken in that paper together with the result quoted above.
Now, we come to the crux of the result. Phase space translations are canonical
transformations \cite{Gol}. Therefore, by a suitable canonical transformation, the Hamiltonian
can be thrown into the following form:
\begin{align}
H(Q,P) = Q_1P_2 + h(Q,P)
\end{align}
Here, $h(Q,P)$ is a function that contains no linear terms. Therefore, in a neighbourhood of the origin, the first term dominates. If we choose the Chaetev's function as $V(Q,P) = Q_1Q_2$ , we get for the derivative of this function along the solution trajectories
$\dot{V} = \dot{Q_1}Q_2 + Q_1\dot{Q_2}$, so that, using the equations of motion contained in (6), 
\begin{align}
\dot{V} = Q_1^2 + u(Q,P)
\end{align}
Again, the function $u(Q,P)$ is dominated by the first term in a neighborhood of the origin.
Therefore, in a small neighborhood around the origin, the conditions of Chaetev's
theorem are satisfied so that this theory is unstable at the origin. Finally, by a suitable
translation, we can reach the same conclusion at any point of the trajectory. Hence, we
conclude that such a theory is unstable \textit{everywhere}.\\
\\
\indent Now that we have shown that any higher derivative Lagrangian is unstable in the
Lyapunov sense, we can return to the question of modifying gravity to explain
accelerated expansion. In the literature,
f(R) modifications are discussed very
extensively, where R is the Ricci scalar [3], [9]. It is claimed that no other non-trivial
modification can satisfy the constraints imposed by the result proved above [3]. However, it turns out (as shown below) that Ostrogradsky's theorem can be overuled for Born-Infeld (B-I) type of modifications to the E-H Lagrangians. 

Lagrangians are particularly attractive because of their
intriguing properties [10].
The B-I Lagrangian first made its appearance with the modification of the
electrodynamic action by Born and Infeld. This was done to remove the infinite self
energy of point charges by introducing an upper bound on the magnitude of the electric
field. It was further pointed out that the B-I electromagnetic field propagates without
birefringence [10]. Analogous work has been carried out for the case of the gravitational
field. The relevant equations of motion can be obtained by varying the action given below :
\begin{align}
S_B = \kappa^{-1} \int \sqrt{det(g_{\mu \nu} + \kappa R_{\mu \nu})}
\end{align}
Two dynamical quantities enter the action in (8). They are the metric and
the connection. The Palatini formalism results if these quantities are considered independent degrees of freedom.
The action is varied with respect to both the metric and
the connection. The more usual method of varying gravitational actions explicitly uses the Levi-Civita connection and treats the metric as the sole dynamical freedom of the gravitational field. Here, we briefly recapitulate (after [5]), the equations of motion resulting from (8) 
\begin{align}
q_{\mu \nu} &= g_{\mu \nu} + \kappa R_{\mu \nu}  \\
\frac{\sqrt |q|}{\sqrt |g|}q^{\mu \nu} &= \lambda g^{\mu \nu} - \kappa T^{\mu \nu}
\end{align}
Note the slight subtlety in notation. $q^{\mu \nu}$ represents the inverse of $q_{\mu \nu}$. $\lambda$ is a cosmological constant which is necessary for consistency. The variation with respect to the connection yields
\begin{align}
\Gamma^{\alpha}_{\beta \gamma} = \frac{1}{2} q^{\alpha \delta} \left(\frac{\partial}{\partial x^{\gamma}}q_{\beta \delta} + \frac{\partial}{\partial x^{\beta}}q_{\gamma \delta}- \frac{\partial}{\partial x^{\delta}} q_{\beta \gamma}\right)
\end{align}

The cosmology based on the Born-Infeld Theory has been restricted to the case where the
spatial hyperspaces are flat [11]. We have extended these results to the case where the
spatial hyperspaces can be curved. Before proceeding further, we must mention here
an important consequence of the Ostrogradsky instability which shows why the Palatini formalism is important for the B-I lagrangian.

\subsection*{The Born Infeld Lagrangian can \textit{only} be varied in the Palatini formalism}
Suppose (8) is varied in the metric formalism.Then as stated in above, the connection is fixed as the Levi
Civita connection. In that case, the determinant in (8), can be expanded in
terms of the trace as follows (upto third order)

\begin{align}
det(g_{\mu \nu} + \kappa R_{\mu \nu}) \approx 1+ \frac{1}{2} (\kappa R-\kappa K- \kappa^{2} S)
\end{align}
where  $K = R_{\mu \nu} R^{\mu \nu} - \frac{1}{2}R^2$ and $S = 8R^{\mu \nu}R_{\mu \alpha}R^{\alpha}_{\nu} - 6RR_{\mu \nu}R^{\mu \nu} + R^3$.\\
Therefore, if the metric is considered the sole dynamical quantity, the Lagrangian would
involve the second and higher derivatives of the metric. But this would result in an
unstable theory in accordance with the result proved above.
\\ \indent However, since only the first derivatives of the connection coefficients
appear in the expression for the Ricci tensor, there is no instability provided the action in (8) is varied by taking the metric and
connection as separate degrees of freedom. Therefore, Born-Infeld Lagrangians can be varied consistently in the Palatini formalism.

\section*{The FRW equations in Born Infeld Gravity}
With the question of viability settled, we
can now set up and solve the FRW equations for B-I gravity. We generalize here the flat
space results of [11] to curved spatial hyper surfaces
As usual, we assume that the universe is homogenous and isotropic. This results in the
following ansatz for the metric in comoving coordinates [2]
\begin{align}
g_{\mu \nu} dx^\mu dx^\nu = -(dx^0)^2 + a(t)^2 \left\{ \frac{(dr)^2}{1-kr^2} + r^2 ((d\theta)^2+sin^2\theta(d\phi)^2) \right\}
\end{align}
For the B-I
equations, we also need an ansatz for the auxiliary metric $q_{\mu \nu}$. This is at once supplied
by the demand for homogeneity and isotropy as [6]
\begin{align}
q_{00} &= - U(t)^2 \\ 
q_{ij} &= V(t)^2 \gamma _{ij}
\end{align}
Here, $\gamma_{ij}$ represents the spatial components of the metric. Equations (14), (15) and (11) then determine the non-zero connection coefficients as 

\begin{align*}
\Gamma^{t}_{tt} &= \frac{\dot{U}}{U}   & \Gamma^{t}_{rr} &= \left\{\frac{a^2 V\dot{V}}{(1-kr^2)U^2} + \frac{a\dot{a}V^2}{(1-kr^2)U^2}\right\}  
\\
\Gamma^{t}_{\theta \theta} &= \left\{\frac{a^2V\dot{V}}{U^2} + \frac{a\dot{a}V^2}{U^2}\right\} & \Gamma^{t}_{\phi \phi} &= \Gamma^{t}_{\theta \theta} sin^2\theta
\\
\Gamma^{r}_{rr} &= \frac{kr}{1-kr^2} & \Gamma^{r}_{\theta \theta} &=  -r(1-kr^2) 
\\
\Gamma^{\theta}_{\phi \phi} &= -sin\theta cos\theta & \Gamma^{\phi}_{\phi \theta} &= cot\theta
\\
\Gamma^{\theta}_{\theta r} &= \Gamma^{\phi}_{\phi r} = \frac{1}{r}& \Gamma^{r}_{\phi \phi} &= \Gamma^{r}_{\theta \theta} \text{sin}^2\theta
\end{align*} 
The $\mu \nu = 00$ and $11$ components of the Ricci tensor are given by 
\begin{align*}
\begin{split}
R_{00} &= 3H_1(H+H_2)-6HH_2-3\left\{\frac{\ddot{a}}{a} + \frac{\ddot{V}}{V}\right\}
\\
R_{11} &= \frac{1}{1-kr^2}\Bigl\{2k+\frac{a^2}{U^2}\Bigl\{2HV\dot{V}+ \dot{V}^2 + V\ddot{V}\\
       &+ V^2\left\{H^2+\frac{\ddot{a}}{a}\right\}+V\dot{V}+HV^2\left\{H+H_1 -H_2 \right\}\Bigr\}\Bigr\}
\end{split}
\end{align*}

\noindent Here, $H = \frac{\dot{a}}{a}$, $H_1 = \frac{\dot{U}}{U}$, $H_2 = \frac{\dot{V}}{V}$

For the matter distribution, we assume a perfect fluid with pressure $p$ and density $\rho$. It
can further be shown that the fluid satisfies the equation of continuity. Finally, we can
write the following equations of motion using (9) and (10)

\begin{align*}
1-U^2 &= \kappa\left\{3H_1(H+H_2)-6HH_2-3\left\{\frac{\ddot{a}}{a} + \frac{\ddot{V}}{V}\right\}\right\}\\
V^2 -1 &= \frac{2k\kappa}{a^2}+\frac{\kappa}{U^2}\Bigl\{2HV\dot{V}+ \dot{V}^2 \\
&+ V\ddot{V}+ V^2\left\{H^2+\frac{\ddot{a}}{a}\right\}+V\dot{V}+HV^2\left\{H+H_1 -H_2 \right\}\Bigr\}\\
UV &= \lambda + \kappa p\\
\frac{V^3}{U} &= \lambda + \kappa \rho\\
\end{align*}
The fluid also has equations of state and continuity  
\begin{align*}
p &= w\rho\\
\frac{\dot{\rho}}{\rho} &= -3H(1+w)
\end{align*}
After doing all the manipulations, we arrive at our final expression for the Hubble rate in B-I cosmology
\begin{align}
H = \frac{\dot{a}}{a} = \frac{\left(\sqrt{\alpha^{2} +\left\{4\left\{\alpha^{2} + 2\beta +2+\alpha \beta - \alpha\right\}\left\{(\frac{U}{V})^2\left(\frac{a^2}{\kappa}(V^2-1)-2k\right) + \frac{(U^2-1)}{3\kappa}\right\} \right\}}-\alpha \right)}{2(\alpha^{2} + 2\beta +2 + \alpha\beta-\alpha)}
\end{align}

Here, $\alpha = -\frac{3\kappa\rho(1+w)}{4}\left\{\frac{w}{\lambda+\kappa w\rho} + \frac{1}{\lambda+w\rho}\right\}$ and $\beta = -\frac{3\kappa\rho(1+w)}{4}\left\{\frac{3w}{\lambda+\kappa w\rho}-\frac{1}{\lambda+w\rho}\right\}$.\\
\\
\begin{figure}
\begin{center}
\includegraphics[scale=0.3]{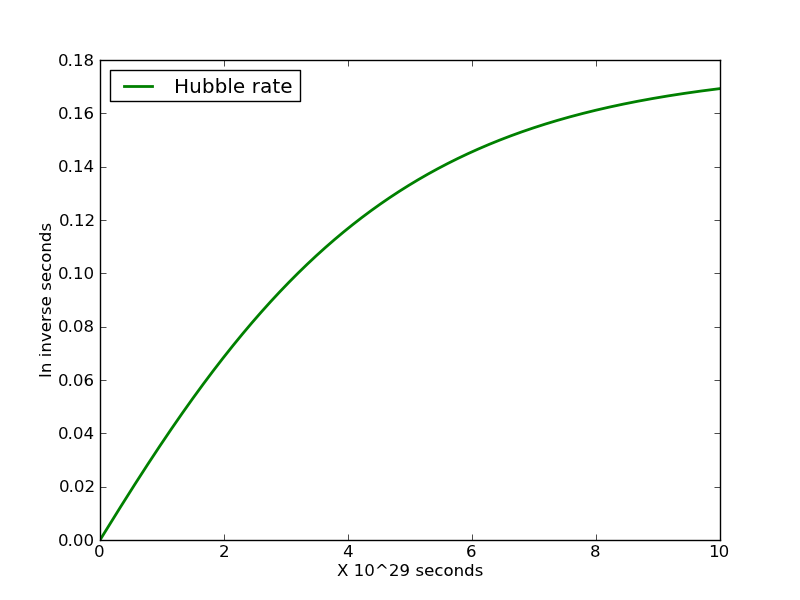}
\caption{The Hubble rate as a function of time for a matter dominated open universe. Present time corresponds approximately to
abscissa = 0.01
}
\end{center}
\end{figure}

\begin{figure}
\begin{minipage}[b]{0.55\linewidth}
\centering
\includegraphics[width=\textwidth]{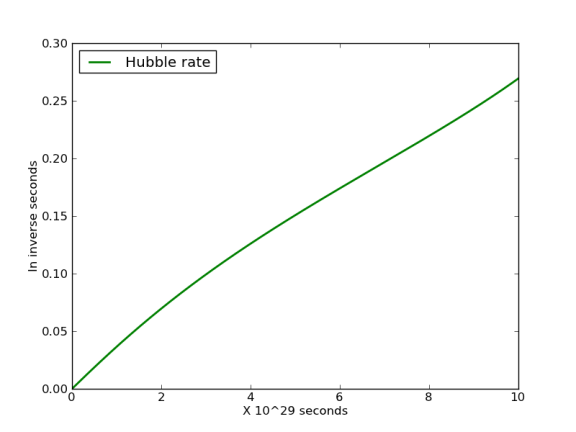}
\caption{The Hubble rate as a function of time for a matter dominated closed universe.
Present time corresponds to abscissa = 0.01. A comparison with the previous figure indicates that the Hubble rate
increases much faster for a closed universe than an open universe. 
}
\end{minipage}
\hspace{0.5cm}
\begin{minipage}[b]{0.55\linewidth}
\centering
\includegraphics[width=\textwidth]{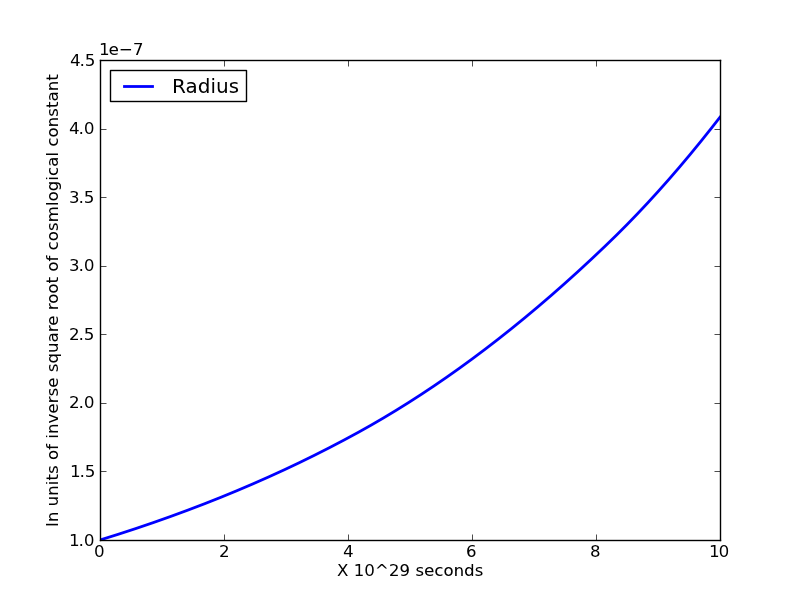}
\caption{The cosmic radius as a function of time for a matter dominated closed universe. As discussed in the text, the acceleration of the
universe is clearly apparent}
\end{minipage}

\end{figure}
\indent As illustrated in Figure 1 and 2, the main point of difference between the solutions of the
FLRW equations in Einstein's gravity and in the present case is that closed universes in
B-I gravity do not experience a 'big-crunch' even after the elapse of a large amount of
time. Moreover, the radius always accelerates in this theory, which can be contrasted
with the case of general relativity, where it always decelerates. An important point that was observed during the plotting of these figures was that the
actual values of the radius as a function of time depended critically on the values of the
coupling constant $\kappa$ and the cosmological constant $\lambda$. Therefore, by tuning these values
properly, one can, in principle, make the actual values of the radius agree with the
experiment. \\
\\
\indent On the other hand, an advantage of f(R) modifications is that any history of
the evolution of the universe can be supported by a proper choice of the function f(R)
[3]. This flexibility has been lost in the case of the Born-Infeld modification.
Nevertheless, as the plots show, the history predicted by our calculations is to
a tolerable extent, consistent with observations.

\textit{Acknowledgements:} We sincerely thank Professor M. Sami for introducing us to Ostrogradsky instability in modified gravity.


\begin{thebibliography}{20}
    \bibitem{Caroles1}S. Carroll et al., Is Cosmic Speed-Up Due to New Gravitational Physics? \textit{Phys. Rev. D} vol.70 Issue 4 Aug. 2004 
    \bibitem{Caroles2}S. Carroll, Lecture Notes on General Relativity, arXiv : gr-qc/ 9712019v1, Dec. 1997
    \bibitem{Wood}R.P. Woodard, Avoiding Dark Matter with 1/R modifications of Gravity, Lecture Notes in Physics, Springer, vol. 720, 2007, pp 403-433
    \bibitem{capoz}S.Capozziello et al., A bird's eye view of f(R) gravity, Open Astronomy Journal, ISSN 1874-3811, Oct. 2009
    \bibitem{Ban}M.Banados and P.G. Ferreira, Eddington's theory of gravity and its progeny, \textit{Phys. Rev. Lett} vol.105, Issue 1, 425-432, Jun. 1979
    \bibitem{Kh}H.Khalil, Non-Linear systems, 2nd Edition, Prentice Hall
    \bibitem{Akk}Ajit Kumar et al., Stability of Charged Solitons, International Journal of Theoretical Physics, vol. 18, Issue 6, pp. 425-432, Jun. 1979
    \bibitem{Gol}Goldstein, Safko, Poole, Classical Mechanics, 3rd Edition
    \bibitem{tsvf}T.Sotiriou and V. Faraoni, f(R) theories of gravity, \textit{Rev. Mod. Phys.} vol.82, Issue 1, Mar. 2010
    \bibitem{DTC}Coulomb Scattering in Born-Infeld electrodynamics \textit{Phys. Rev. D} vol.83, Issue 4, Feb. 2011
    \bibitem{Ban2}M.Banados, P.G. Ferreira and C.Skords, Eddington-Born-Infeld gravity and the large scale structure of the universe, \textit{Phys. Rev. D.} vol.79, Issue 6, Mar. 2009
    \bibitem{AJK}S. Perlmutter, \textit{Phys. Today} 56, 4, 53 (2003)
\end{thebibliography}
\end{document}